\documentclass[12pt]{article}
\usepackage{latexsym}
\usepackage{geometry} 
\usepackage{pict2e}

\def\bs{\begin{subequations}}
\def\es{\end{subequations}}

\catcode`\@=11

\newtoks\@stequation
\def\subequations{\refstepcounter{equation}
  \edef\@savedequation{\the\c@equation}%
  \@stequation=\expandafter{\theequation}
  \edef\@savedtheequation{\the\@stequation}
  \edef\oldtheequation{\theequation}%
  \setcounter{equation}{0}%
  \def\theequation{\oldtheequation\alph{equation}}}

\def\endsubequations{\setcounter{equation}{\@savedequation}%
  \@stequation=\expandafter{\@savedtheequation}%
  \edef\theequation{\the\@stequation}\global\@ignoretrue}

\makeatletter
        \renewcommand{\theequation}{\thesection.\arabic{equation}}%
        \@addtoreset{equation}{section}%
\makeatother

\renewcommand{\thefootnote}{\fnsymbol{footnote}}

\begin{document}

\begin{titlepage}
 June 15, 2019   - revision accepted for publication in Int. J. Mod. Phys. A
\begin{center}        \hfill   \\
            \hfill     \\
                                \hfill   \\

\vskip .25in

{\large \bf An Approach for Modelling Tachyons with Gravitation \\}

\vskip 0.3in

Charles Schwartz\footnote{E-mail: schwartz@physics.berkeley.edu}

\vskip 0.15in

{\em Department of Physics,
     University of California\\
     Berkeley, California 94720}
        
\end{center}


\vfill

\begin{abstract}
This work expands previous efforts, within the classical theories of Special and General Relativity, to include tachyons (faster-than-light particles) along with ordinary (slower-than-light) particles at any energy. The objective here is to construct a Hamiltonian that includes both the particles and the gravitational field that they produce. We do this with a linear approximation for the Einstein field equations; and we also assume a time-independent gravitational metric implied by a static picture of the particles' motion. The resulting formulas will allow serious modelling to test the idea that cosmic background neutrinos may be tachyons, which can produce the observed gravitational effects now ascribed to some mysterious Dark Matter.

\end{abstract}

\vfill

\end{titlepage}

\renewcommand{\thefootnote}{\arabic{footnote}}
\setcounter{footnote}{0}
\renewcommand{\thepage}{\arabic{page}}
\setcounter{page}{1}

\section{Introduction}
Some readers may wish to start with Appendix A.

In earlier work  \cite{CS1, CS2, CS3, CS4} I have explored, theoretically, how tachyons (faster-than-light particles) would behave in Einstein's General Relativity. That starts with the recognition that low energy tachyons would have very large velocities and thus their contribution to the energy-momentum tensor $T^{ij}$ would be very large. That leads to the physical idea that neutrinos, which are so numerous throughout the universe, might be tachyons with a mass of around 0.1 eV and could thus produce gravitational effects that are now ascribed to mysterious sources called "Dark Matter" or "Dark Energy".

The simplest calculation, based upon an unexpected minus sign in front of $T^{ij}$, gave a numerical estimate of the negative pressure that such tachyon-neutrinos would produce in the Robertson-Walker model of the universe that "explained" Dark Energy within a factor of 2! \cite{CS3, CS4}

The more difficult calculation surrounds the idea that attractive gravitational forces among low energy tachyons could lead to their forming stable configurations (while all the time flowing at speeds far above the speed of light) that could attach themselves to galaxies and thus produce the local gravitational fields that are commonly ascribed to Dark Matter.\cite{CS1} This paper is about that challenging idea.

For a non-relativistic classical particle moving in a time-independent conserved force field we have the familiar equation, expressing the conservation of energy:
\begin{equation}
E = \frac{1}{2} mv^2 + V(x), \label{z1}
\end{equation}
and for a collection of  such particles interacting via Newtonian gravity we write the Hamiltonian,
\begin{equation}
H = \sum_a \frac{1}{2} m_a v_a^2 - \sum_{a < b} \frac{Gm_a m_b}{r_{ab}}, \;\;\;\;\; r_{ab} = |\textbf{x}_a - \textbf{x}_b|.\label{z2}
\end{equation}

The main purpose of this paper is to generalize this Hamiltonian to the case of relativistic matter, including both ordinary particles ($v < c=1$) and tachyons ($v > c=1$), under Einstein's General Theory of Relativity, with two restrictions: that we use the linearized approximation to Einstein's equation; and that we assume the metric $g_{\mu \nu}(x)$ to be independent of the time.  The result given in Section 7, for a system of ordinary particles and tachyons at low (or intermediate) energies, is,
\begin{eqnarray}
\textbf{Formula 1}:\;\;\;H = -\sum_a \omega_a  E_a   - \sum_{a , b} \frac{G \;\omega_a E_a  \; \omega_b E_b }{r_{ab}} Z_{ab},\;\;\;\;\;\;\;\;\;\; \nonumber \\
Z_{ab} = 2 -4 \textbf{v}_a \cdot \textbf{v}_b + v_a^2 + v_b^2 -[\epsilon_a \gamma_a^2 + \epsilon_b \gamma_b^2 +1] [(1- \textbf{v}_a \cdot \textbf{v}_b )^2 - (1/2)(1-v_a^2)(1-v_b^2)]. \label{z3}
\end{eqnarray}
Here  $E= \sqrt{p^2 \pm m^2} = m\gamma = m/\sqrt{|1-v^2|}$; and the factors $\epsilon, \omega$ are $\pm1$ and will be defined later. The expression $Z_{ab}$ contain all the details of the velocity-dependent interactions.

For other energy ranges we also give two additional formulas.

The idea of the metric being static while the particles producing that gravitational field are moving may seem contradictory. We imagine a continuous flow of particles that does not change in time, analogous to the picture of a constant electrical current used in the study of magnetostatics. One might use the term Gravito-Statics for this study. (But that name has been used for a different sort of theory: see \cite{PR}). 

There is also mathematical work on the static Einstein-Vlasov system \cite{HA} which uses a kinetic theory approach for many-particle systems in General Relativity; but that does not consider the possibility of tachyons. Therefore I shall begin from scratch.

Moreover, the large body of work on the Einstein-Vlasov equations (see references in \cite{HA}) is focused on the equations of motion and the mathematical question of the existence of solutions. That work has not been productive in looking at the stability of such solutions. By contrast, my Hamiltonian formalism leads readily to such analyses from a physical perspective; and in the latter part of Section 7 I do draw a number of provocative suggestions about how these models may lead to new physical insights. In particular, these new equations lead me to discard, as clearly unstable, my original notion \cite{CS1} of tachyons grouping into long ropes held  together by their mutual gravity.

With this formula (\ref{z3}) we can begin model-building, looking for potentially stable configurations of tachyon flows contained by their mutual gravity. That will be an ongoing task. 
That requirement of stability will be the most challenging. Even with ordinary Newtonian gravity (\ref{z2}) one sees large scale attraction that seems to lead inevitably to physical collapse; but then one brings in further physics to help us explain the observable stability of stars, solar systems, galaxies. Our new ideas are about incorporating tachyons (neutrinos?) into that cosmic modelling; and it needs to start with something better than (\ref{z2}): thus our new formula (\ref{z3}).

\section{Beginning} 

Here I review previous work describing both ordinary particles ($v < c$) and tachyons ($v>c$)  as classical particles in both Special and General Relativity.  First, some notation and equations in common for all particles.

A "world line" $\xi^\mu (\tau) = (t(\tau), \textbf{x}(\tau))$ maps the trajectory of the particle in space and time with some as yet undefined scalar parameter $\tau$. I use notation $\mu = (0,i), i = 1,2,3$ and set the velocity of light $c=1$; and use the overhead dot notation to represent $d/d\tau$. The argument x is meant to stand for all four spacetime coordinates $x^\mu = (t,\textbf{x})$, where $\textbf{x} = (x^1, x^2, x^3)$; and partial derivatives are written as $\partial_\mu = \frac{\partial}{\partial x^\mu}$.

We write a conserved current density as 
\begin{eqnarray}
j^\mu (x) = \int d\tau\; \dot{\xi}^\mu(\tau)\; \delta^4 (x -\xi(\tau)), \label{a1} \\
j^\mu_{\;,\mu}(x) = \partial_\mu j^\mu(x) = \int d\tau\; \dot{\xi}^\mu(\tau) \partial_\mu\; \delta^4(x - \xi(\tau)) = \label{a2}\\
- \int d\tau\; \frac{d \xi^\mu}{d\tau} \frac{d}{d\xi^\mu}\;\delta^4 (x - \xi (\tau)) = 
- \int d\tau \frac{d}{d\tau}\;\delta^4 (x - \xi (\tau)) = 0.\label{a3}
\end{eqnarray}
All that we required for that last step was to take the end points of the $\tau$ integral far away from the place where the particle is at this location x.

We can also write an energy-momentum tensor,
\begin{equation}
T^{\mu \nu} (x) = m \int d\tau \; \dot{\xi}^\mu(\tau)\; \dot{\xi}^\nu(\tau)\; \delta^4 (x - \xi(\tau)). \label{a4}
\end{equation}
When we take the divergence on one index we follow the above calculation but get something left over from the final partial integration.
\begin{equation}
T^{\mu \nu}_{\;\;\;,\mu} (x) = -m\int d\tau\dot{\xi}^\nu \frac{d}{d\tau}\;\delta^4(x-\xi(\tau)) =m \int d\tau \ddot{\xi}^\nu \; \delta^4(x-\xi(\tau)). \label{a5}
\end{equation}

If the only forces acting upon the particle are those due to gravity, then we also have the geodesic equation (equally correct for ordinary particles or tachyons):
\begin{equation}
\ddot{\xi}^\nu + \Gamma^\nu _{\alpha \beta} \;\dot{\xi}^\alpha\;\dot{\xi}^\beta = 0,   \label{a6}
\end{equation}
involving the Christoffel symbols $\Gamma^{\nu}_{\alpha \beta}$, defined in terms of derivatives of the metric tensor $g_{\mu \nu} (x)$,  evaluated at the point where the particle is at any given value of $\tau$. This lets us write the result of the ordinary divergence calculation as,
\begin{equation}
T^{\mu \nu}_{\;\;\;,\mu} + \Gamma^\nu_{\alpha \beta}T^{\alpha \beta} = 0. \label{a7}
\end{equation}
From this result we can construct a modified tensor, multiplied by the square root of the determinant of the metric, which will have zero as its covariant divergence. 
\begin{equation}
{\cal{T}}^{\mu \nu} = \sqrt{|det(g)|} T^{\mu \nu}, \;\;\;\;\; {\cal{T}}^{\mu \nu}_{\;\;\; ;\mu} = 
{\cal{T}}^{\mu \nu}_{\;\;\; ,\mu}  + \Gamma^\mu_{\alpha \mu}{\cal{T}}^{\nu \alpha}  + \Gamma^\nu_{\alpha \mu}{\cal{T}}^{\alpha \mu} =0. \label{a8}
\end{equation}

That is proper for the full Einstein equation; but here we will be satisfied with the linear approximation.
\begin{eqnarray}
g_{\mu \nu} = \eta_{\mu \nu} + h_{\mu \nu} - \frac{1}{2} \eta_{\mu \nu} h, \;\;\; h = \eta^{\mu \nu} h_{\mu \nu},\;\;\;\partial^\mu h_{\mu \nu} = 0, \label{a9}\\
\partial^\alpha \partial_\alpha h_{\mu \nu}(x)= [\partial_t^2 - \nabla \cdot \nabla]h_{\mu \nu}(x) = -16\pi G T_{\mu \nu}(x) \label{a10},
\end{eqnarray}
with the Minkowski metric $\eta_{\mu \nu}=\delta_{\mu \nu} (+1, -1, -1, -1)$ used to raise and lower indices.

\section{Free particles}

First, we sit in a flat spacetime. There should be nothing new here; we want to practice for later. At any point in the (ordinary) particle's trajectory, where it happens to have a velocity $v$ in the original reference frame we can make a Lorentz transformation to a local frame where the particle is seen momentarily at rest (in the rest frame: $v\prime=0$). This Lorentz transformation has a velocity $v_{LT} = v$ and the gamma factor $\gamma_{LT} = 1/\sqrt{1-v^2}$. We take the scalar parameter for this particle to be the time in that local rest frame: $d\tau = dt' =dt /\gamma_{LT}$. This $\gamma_{LT}$ is exactly the $\gamma$ of the particle at that point of its trajectory in the original frame. Thus we can write,
\begin{equation}
\dot{\xi}^\mu = (\gamma, \gamma \textbf{v}), \;\;\; \textbf{v} = \frac{d \textbf{x}}{dt}. \label{b1}
\end{equation}

This leads to, with the Minkowski metric,
\begin{equation}
\dot{\xi}^\mu \dot{\xi}^\nu \eta_{\mu \nu} = \gamma^2 (1-v^2) = +1,  \label{b2}
\end{equation}
which is fine for a free particle; but in a gravitational field it should be the entire metric $g_{\mu \nu}$ that fills this role. The geodesic equation implies that $\dot{\xi}^\mu \dot{\xi}^\nu g_{\mu \nu}$ is a constant along the particle's trajectory.

But let's stay in Minkowski space, no gravity, for a while and look at the comparable calculation for tachyons.  Now we have no rest frame to give us a nice definition of the scalar parameter $\tau$. So we find another special frame of reference: one where the tachyon has infinite velocity. (This is the one used in quantum group theory to find the "Little Group" for tachyons.) This involves a velocity of the Lorentz transformation, $v_{LT} = 1/v$, where this $v$ is the tachyon's velocity ($v > 1$) in the original reference frame. (To see this, recall the velocity addition formula $v' = (v+v_{LT})/(1-v v_{LT})$.) The gamma for this Lorentz transformation is $\gamma_{LT} = 1/\sqrt{(1-v^{-2})}$. We now make the definition of the scalar parameter as $d\tau = \hat{v} \cdot d\textbf{x}'$, marking the path of the infinitely fast particle. Here $d\textbf{x}'$ is the differential of the spatial coordinate in this special reference frame, which is related to that in the original frame by $\hat{v} \cdot d\textbf{x}' = \gamma_{LT} ^{-1} \hat{v} \cdot d\textbf{x}$. So we have, for tachyons, in the original reference frame,
\begin{equation}
d\tau = (v\gamma)^{-1}\; \hat{v}\cdot d\textbf{x}, \;\;\; \gamma = 1/\sqrt{(v^2 -1)}, \;\;\; \dot{\xi}^\mu = \gamma  (1, \textbf{v}),  \label{b3}
\end{equation}
where that last equation is what we expected.

For ordinary particles of extremely low velocities, we had $d\tau \approx dt$; for tachyons of extremely high velocities we have $d\tau \approx \hat{v} \cdot d\textbf{x}$. This is nice. 

Now let's look at how these results influence our physical interpretation of the conserved currents.
For ordinary particles,
\begin{equation}
j^\mu = \int d\tau (\gamma, \gamma \textbf{v}) \delta(t-\gamma\tau) \delta^3(\textbf{x} - \gamma\textbf{v} \tau) = 
(\gamma, \gamma \textbf{v}) \gamma^{-1} \delta^3(\textbf{x} - \textbf{v} t),  \label{b4}
\end{equation}
where I have used the first delta-function to do the integral over $\tau$. If we now do the standard integral over all space, we get a very simple answer,
\begin{equation}
\int d^3 x j^0 = 1.  \label{b5}
\end{equation}
This says we have one particle there (somewhere in space) at any time.

The analogous calculation for a tachyon goes like this.
\begin{equation}
j^\mu = \int d\tau \gamma (1,  \textbf{v}) \delta(t-\gamma  \tau) \delta^2(x_\perp)\delta(x_\parallel - \gamma  v \tau),  \label{b6}
\end{equation}
where the parallel and perpendicular subscripts refer to the direction of the velocity. Now I do the integration over $\tau$ using the last delta-function to get, 
\begin{equation}
j^\mu = (1, \textbf{v})\; \frac{1}{v}\;\delta (t - x_\parallel / v) \delta^2 (x_\perp).  \label{b7}
\end{equation}
To get the count of "one particle" from this I integrate the component of $\textbf{j}$ parallel to the velocity over the transverse plane and integrate over time.
\begin{equation}
\int dt \int d^2 x_\perp \hat{v} \cdot \textbf{j} = 1.  \label{b8}
\end{equation}
We recite this conservation law as: We have one particle passing through a transverse plane at some time - and this could be any transverse plane. This is completely consonant with my earlier writings about tachyon kinematics.

Suppose I try to treat the tachyon as I did the ordinary particle. I still go to the frame where the tachyon is at $v'=\infty$ but I choose to define $d\tau = dt'$ in that frame. I again write $dt' = dt/\gamma_{LT}$ but remember that $\gamma_{LT} =1/\sqrt{(1-1/v^2)} = v \gamma$, where this last is the usual gamma for the tachyon in the original reference frame. This gives us $\dot{\xi}^\mu = \gamma v (1,\textbf{v})$. We now calculate the current, as before, 
\begin{equation}
j^\mu = \int d\tau v\gamma (1, \textbf{v}) \delta(t-v\gamma\tau) \delta^3(\textbf{x} - v\gamma\textbf{v} \tau) = (1, \textbf{v}) \delta^3(\textbf{x} - \textbf{v} t),  \label{b9}
\end{equation}
integrating over $\tau$ by using the first delta-function. This looks just like the case with ordinary particles. We are tempted to integrate over $d^3 x$ and say that we have one particle in a large box at any time - just as we did for ordinary particles. However, this is really not acceptable for tachyons: the velocity $v$ occurs in that delta-function $\delta (\textbf{x} - \textbf{v}t)$ and that velocity can be arbitrarily large. Thus, given any finite "box" over which we do the $\int d^3x$ at time $t_1$ there may be tachyons that will be located out of that box at time $t_2$. (This cannot happen for ordinary particles.) We conclude, as in earlier writings, that the first method of treating tachyons - using space displacement to define the parameter $\tau$ - is the correct one for them.

What if we take this second version of $j^\mu$ for tachyons and integrate it as we did for the first version.
\begin{equation}
\int dt \int d^2 x_\perp \hat{v} \cdot \textbf{j} = v \int dt \delta (x_\parallel - vt) = 1,  \label{b10}
\end{equation}
which looks nice. 

We can apply the same analysis to the energy-momentum tensor: just add the factor $m\dot{\xi}^\nu$ to the results above for $j^\mu$. 
\begin{eqnarray}
T^{\mu \nu} = m \gamma (1, \textbf{v}) (1, \textbf{v})\;\delta^3 (\textbf{x} - \textbf{v} t),  \label{b11} \\
or \nonumber \\
T^{\mu \nu} = m \gamma  (1, \textbf{v}) (1, \textbf{v})\;v^{-1} \; \delta(t - x_\parallel/v) \delta^2 (x_\perp),  \label{b12}
\end{eqnarray}
for the ordinary particle or tachyon, respectively. 

\section{Lagrangian/Hamiltonian formalism}

Textbooks show how to write a Lagrangian and a Hamiltonian for a single relativistic (ordinary) particle; but then say that one cannot make this "manifestly covariant" for many-particle systems because each individual particle's $d\tau$ is independent of the others'. Traditional Lagrangian formalism involves particles and fields but all described on a common space-time manifold. When we come to add gravitation, there is the familiar caveat that "Energy" is not well defined in Einstein's theory, at least because there are possible gravitational waves that need to be attended to; but that need not bother us here. There is also a sophisticated literature about the "positive energy theorem" in general relativity \cite{EW}; but that work explicitly requires $T_{00} \ge 0$ everywhere in each local Lorentz frame and that would prohibit our inclusion of tachyons. 

Our objective here is to study a "static" physical system of particles - both ordinary and tachyons -  with gravitational interaction, derived from Einstein's equation. By the word "static" we mean that the particles are moving, but their pattern of flow does not change with time. This should imply that the gravitational field they produce - via the metric $g_{\mu \nu} (x)$ - is independent of the time. But this must mean that we are restricting ourselves to one (or a particular set of) Lorentz frames. If any field is independent of time (but varying with spatial position) in one reference frame, a Lorentz transformation that takes us to a frame moving relative to the original frame will show the field (at any place) as varying with time. So, our final analysis will be done in a particular Lorentz frame: and this is ok. Nevertheless we want to start with a generally invariant/covariant formalism, and specialize to the static case later.

I want to be especially careful about minus signs here. Start with one particle:

\begin{eqnarray}
L = \int d^3x\; {\cal{L}}, \;\;\; {\cal{L}} = \zeta m \int d\tau\; \sqrt{\dot{\xi}^\mu(\tau) \dot{\xi}^\nu (\tau)\;\epsilon \eta_{\mu \nu}} \;\delta^4(x - \xi (\tau)),  \label{c1}\\ 
L = \zeta m \int d\tau\; \sqrt{\dot{\xi}^\mu(\tau) \dot{\xi}^\nu (\tau)\;\epsilon \eta_{\mu \nu}} \;\delta(t- \xi ^0(\tau)). \label{c2}
\end{eqnarray}
Here the dot means derivative wih respect to $\tau$, $\eta_{\mu \nu}$ is the Minkowski metric; $\epsilon = \pm 1$ distinguishes ordinary particles from tachyons; and $\zeta$ is another $\pm 1$ factor that we will have to argue about later on. We now use the remaining delta-function to eliminate the integral over $\tau$ - and this leaves us with a factor $|\dot{\xi}^0| ^{-1}$. We now write,
\begin{equation}
\dot{\xi}^\mu = (dt/d\tau, d\textbf{x}/d\tau) = \dot{\xi}^0 (1, \textbf{v}), \;\;\; \textbf{v} = d\textbf{x}/dt. \label{c3}
\end{equation}
and this yields,
\begin{equation}
L = \zeta m \;\sqrt{\epsilon(1-v^2)}. \label{c4}
\end{equation}
This is for all species of particles, ordinary or tachyon.

We then proceed with the "canonical" formalism,
\begin{equation}
p_i = \frac{\partial L}{\partial v^i} = \zeta\epsilon m (-v^i)\gamma, \;\;\; \gamma = 1/\sqrt{\epsilon(1-v^2)}, \;\;\;
H = p_i v^i -L = - \zeta \epsilon m \gamma. \label{c5}
\end{equation}
For ordinary particles ($\epsilon = +1$) at low velocities this gives,
\begin{equation}
H = -\zeta (mc^2 + 1/2 mv^2 + ...) \label{c6}
\end{equation}
Thus it is conventional to choose $\zeta = -1$. For tachyons ($\epsilon = -1$) we have,
\begin{equation}
H = +\zeta m \gamma. \label{c7}
\end{equation}
It is tempting to choose $\zeta = +1$ but maybe we should wait to see about this sign.

The final step in going from Lagrangian to Hamiltonian is to eliminate the velocity variable v in favor of the momentum variable p. From Eq. (\ref{c5}) we calculate,
\begin{equation}
 p^2 \equiv \sum_i (p_i)^2 = m^2 v^2 \gamma^2, \;\;\;\;\; H =- \zeta \epsilon m \gamma = - \zeta \epsilon \sqrt{p^2 + \epsilon m^2}, \label{c7a}
\end{equation}
which looks very familiar.

For many particles, labelled with the subscript "a", we now write the Lagrangian density for all this matter, in the presence of a gravitational field as follows.
\begin{equation}
{\cal{L}}_M (x)= \sum_a\;  \zeta_a m_a \int d\tau_a\; \sqrt{\dot{\xi}_a^\mu(\tau_a) \dot{\xi}_a^\nu (\tau_a)\;\epsilon_a \; g_{\mu \nu}(x)} \;\delta^4(x - \xi _a(\tau_a)), \label{c8}
\end{equation}
and, following the method used above for each particle's coordinates,
\begin{equation}
L_M = \sum_a \zeta_a m_a \;\sqrt{\epsilon_a g_{\mu \nu} (\textbf{x}_a) v_a^\mu v_a ^\nu}, \label{c9}
\end{equation}
where $v^\mu_a = (1, \textbf{v}_a) = (1, d\textbf{x}_a/dt)$. What we have here, for the physical problem posed in Section 1, is an expression where the metric $g_{\mu \nu}$ does not depend explicitly on the time t; it does depend on the coordinates of the particle $\textbf{x}_a$ in each term, and those coordinates do depend on the time t. The particle velocities $\textbf{v}_a$ also depend implicitly on the time t. So we can do conventional steps of Lagrangian analysis, as follows.
\begin{equation}
p_{a\;\mu} = \frac{\partial L_M}{\partial v^\mu_a} = \zeta_a m_a \epsilon_a g_{\mu \nu}(\textbf{x}_a) v_a^\nu/ \sqrt{\epsilon_a g_{\mu \nu} (\textbf{x}_a) v_a^\mu v_a ^\nu}. \label{c10}
\end{equation}
Since we have defined $v^0 = 1$ this equation should be read only for $\mu = i = 1,2,3$ in terms of Lagrangian formalism. However, as we shall see below, this may be read as a generally covariant definition of momentum.
We also have the geodesic equation for each individual particle,
\begin{equation}
\ddot{\xi}_a ^\mu + \Gamma^\mu _{\alpha \beta} \dot{\xi}_a^\alpha \dot{\xi}_a^\beta = 0, \label{c11}
\end{equation}
which comes from varying each worldline $\xi_a (\tau_a)$ in the action made with this Lagrangian density (\ref{c8}) in the most general case. (This geodesic equation does not involve the factors $\zeta , \epsilon$.)From this geodesic equation we have, in the general case, the integral,
\begin{equation}
g_{\mu \nu} (x = \xi_a(\tau_a)) \dot{\xi}_a ^\mu \dot{\xi}_a ^\nu = constant = \epsilon_a \kappa_a^2 \label{c12}
\end{equation}
Here I put in the factor $\epsilon_a = \pm 1$ to distinguish the two species of particles we study; and I also put in some constants $\kappa_a$. As usual, we make these constants $\kappa_a$ equal to 1 by scaling the previously arbitrary parameters $\tau_a$.

With the above information, we now calculate the Hamiltonian for these particles,
\begin{equation}
H_M = \sum_a p_{a\;i} v_a^i -L_M = - \sum_a p_{a \; 0}. \label{c12a}
\end{equation}
For any situation where the metric $g$ is independent of the time, there is a textbook proof \cite{BS1} that the time component of the covariant momentum $p_{a\;0}$ is constant along the particle's worldline. So what we have here is independent of the time.

For any individual particle, using the definition $p^\mu = m\dot{\xi}^\mu$  together with previous formulas in this section, we can write $p_{\mu} = \zeta \epsilon g_{\mu \nu} p^\nu$. Except for the factors $\zeta, \epsilon$ this formula conforms with the standard relation between "covariant" and "contravariant" 4-vectors.

The next task is to rewrite the general formula (\ref{c12a}) in a more useful way. We do this in the following Section.

\section{Expanding the particle Hamiltonian}

In non-relativistic physics we write the Hamiltonian as $H = K.E. + P.E.$ and say that it is time independent.  In General Relativity, as treated in this study, we do have an expression for the Hamiltonian, a constant of the motion, but we need to figure out how to separate it into those two parts called Kinetic Energy and Potential Energy. The linear approximation to Einstein's equations for the metric will be our guide.

We will write $g_{\mu \nu} = \eta_{\mu \nu} + \lambda_{\mu \nu}$, where $\eta$ is of order zero in the gravitational constant G and $\lambda$ is first order in G. It would seem that our task is merely to expand the particle Hamiltonian, given in Section 4, in this same way: the K.E. part will be zero order in G (i.e., the energy of a free particle) and the P.E. part will be everything first order in G. But is this a clear definition? There are other variables that occur: there are spatial coordinates $\textbf{x}$ and velocities $\textbf{v}$ and momenta $p^i$ and $p_i$. How are these to be grouped?

We find, below, that seeking this expansion in G may take different paths, depending on the range of energies (velocities) of the particle.  

We have the defining equation, for one particle of any species,
\begin{eqnarray}
p_0 = \zeta \epsilon m (g_{00} + g_{0i}v^i)\Gamma, \;\;\;\;\;  \Gamma \equiv [\epsilon v^\mu v^\nu g_{\mu \nu}]^{-1/2},\label{cc1}\\
\Gamma^2 = \epsilon/[1-v^2 + \Lambda], \;\;\; \Lambda \equiv \lambda _{00} + 2\lambda_{0i}v^i + \lambda_{ij}v^iv^j.\label{cc2}
\end{eqnarray}
This $\Gamma$ is not the Christoffel symbol. For $G = 0$, $\Gamma = \gamma = 1/\sqrt{\epsilon(1-v^2)}$. 

Now I am ready to make an expansion, since $\Lambda \sim O(G)$. But this requires me to say something about the magnitude of $\gamma$. For what I will call Formula 1, it is assumed that $\gamma$ is not much bigger than 1; that means that $v$ is not close to 1.
\begin{eqnarray}
\textbf{Formula 1}:\;\;\;\Gamma \approx \gamma [1 - (1/2)\epsilon \gamma^2 \Lambda], \\
p_0 \approx \zeta \epsilon m \gamma[1 + \lambda_{00} + \lambda_{0i} v^i - (1/2) \epsilon \gamma^2 \Lambda].\label{cc3}
\end{eqnarray}
This is the formula I gave in the first version of this paper\cite{CS5}. It should be good for ordinary particles and for tachyons at low energies and also perhaps for energies up to some modest multiple of their mass. In that earlier paper I was incautious about using this formula and made some ill-considered attempt to interpret it at high energies. If one looks at the expression (\ref{cc2}) for $\Gamma$ one sees that there is some very bad behavior as one approaches $v=1$: the function is not analytic because of the factor $\epsilon$, which changes from +1 to -1 as one crosses from $v<1$ to $v>1$. Now we know to use this formula only for low energy particles.

For many particles, we put subscripts "a" on all the particle variables and sum them. In the non-relativistic limit for ordinary particles this becomes,
\begin{equation}
H_M \rightarrow \sum_a m_a \gamma_a [1 + (1/2) \lambda_{00}] = \sum_a m_a (1+(1/2)v_a^2)  - \sum_{a, b} \frac{G m_a m_b}{r_{ab}},\label{cc4}
\end{equation}
where I have taken the formula for $\lambda_{00}$ derived in the next Section. 
This looks nice; but it is not quite right. The potential energy for this system should count each pair of particles only once and this formula has an excess factor of 2 in the P.E. This will be corrected when we add in the Hamiltonian for the field.

Another way of characterizing the above approach is to say that we have used only the velocities, and not the momenta, in identifying the K.E. part.  Now we explore the alternatives.

I want to look at the momentum variables in the fully relativistic formulation. We have, in general, the metric $g_{\mu \nu}$ providing the relation between the covariant and contravariant forms, along with the overall constant of motion derived from the geodesic equations:
\begin{equation}
p_\mu = g_{\mu \nu} p^\nu, \;\;\;  g_{\mu \nu}p^\mu p^\nu  = \epsilon m^2. \label{cc5}
\end{equation}
where I have temporarily dropped the phase factor $\omega = \zeta \epsilon$ seen earlier. In order to further simplify this examination, I will also take the metric to be strictly static: $g_{0i} = 0.$\cite{BS2}

I can now write,
\begin{eqnarray}
g_{00} (p^0)^2  + g_{ij} p^i p^j = \epsilon m^2,\;\;\; p_0 = g_{00} p^0,  \\
p_0 = \sqrt{[\epsilon m^2 - g_{ij}p^i p^j]g_{00}}.\label{cc6}
\end{eqnarray}
Next we start the expansion of the metric $g = \eta + \lambda + O(G^2)$
where $\eta$ is the Minkowski metric. This leads to
\begin{equation}
p_0 = \sqrt{[(E^*)^2 - \lambda_{ij}p^i p^j](1+\lambda_{00})}, \;\;\; E^* \equiv \sqrt{p^i p^i + \epsilon m^2} \label{cc7}
\end{equation}

An alternative is to start with the covariant momenta,
\begin{eqnarray}
g^{\mu \nu} p_\mu p_\nu = \epsilon m^2, \\
g^{00} (p_0)^2 +g^{ij} p_i p_j = \epsilon m^2, \\
p_0 = \sqrt{[\epsilon m^2 - g^{ij} p_i p_j ]/g^{00}}\label{cc8}
\end{eqnarray}

Since the matrix $g^{\mu \nu}$ is the inverse of the matrix $g_{\mu \nu}$ we have to first order (and with the simplification of $g_{0i} = 0$): $g^{00} = 1-\lambda_{00}$, $g^{ij} = -\delta_{ij} - \lambda_{ij}$.  This leads to
\begin{equation}
p_0 = \sqrt{(E_*^2 + \lambda_{ij}p_ip_j)/(1-\lambda_{00})}, \;\;\; E_* \equiv \sqrt{p_ip_i +\epsilon m^2}. \label{cc9}
\end{equation}

Comparing the two equations (\ref{cc7}) and (\ref{cc9}) we see a sharp difference: the sign of the $\lambda_{ij}$ term is changed.  If we make expansion to first order in $\lambda$, we see two equations that look rather different. We take these as Formula 2 and Formula 3.
\begin{eqnarray}
&&\textbf{Formula 2}: \;\;\;p_0 \approx E^*  + E[\lambda_{00} - \lambda_{ij} v^i v^j]/2, \label{cc10}\\
&&\textbf{Formula 3}: \;\;\;p_0 \approx E_* +  E[\lambda_{00} + \lambda_{ij} v^i v^j]/2.\label{cc11}
\end{eqnarray}
Here I have set $E^* = E_*=E$ when multiplying the (first order small) quantity $\lambda$; and I also wrote $p/E=v$ for both cases in the same circumstance.
These two formulas are two different ways of expressing the same quantity, $p_0$. Formula 2 is exactly  what I put forward as the primary solution in the second version of this paper \cite{CS5}. My error there was to also look at this formula for low energy tachyons; but now we see that these two formulas should \emph{not} be used for low energy tachyons ($E \rightarrow 0$).

Let's remember, the momenta involved in these two expressions are different and thus $E^*$ and $E_*$  are also different.  Both the covariant and the contravariant momenta depend on the gravitational field: $p^i = m \Gamma v^i$, and  $p_i = -p^i + \lambda_{ij}p^j$. Thus $p_i p_i = p^i p^i - 2 \lambda_{ij}p^ip^j$; and we see that the two formulas for $p_0$ are actually identical. But this still leaves us with the challenge of deciding which one to use.  At least we can say that these last 2 formulas should \emph{not} be used for low energy tachyons; this is because $E \rightarrow 0$ would prevent us from expanding the above square root expressions in the way we have done.
In the case of low energy ordinary particles $v \rightarrow 0$, these two formulas are the same; and in fact they are the same as Formula 1.

Our goal was to separate this particle Hamiltonian, which is given by $p_0$,  into a K.E. part and a P.E. part. We wanted to say that the K.E. term is zeroth order in G and the P.E.  term (involving $\lambda$) is first order in G. Thus, which formula we start with is an important choice as we expand to first order in G. However, we now recognize that introducing the momenta and using them to define the free particle energy E  (as $E^*$ or $E_*$ ) will involve some G dependence in the K.E. term.This also incorporates the spatial coordinates $\textbf{x}$ in the K.E. Is this bad or not?

Canonical formalism (Lagrangian/Hamiltonian) would lead us to use momentum variables throughout the Hamiltonian; in particular we have noted that the covariant momenta are the canonical variables for our system. Is this a rule we need to follow?

Given these open questions, we nevertheless proceed to see what we have.

In the following Sections we shall first derive formulas for the $\lambda$'s and then go on to assemble the complete Hamiltonian, for particles and for the gravitational field,  under each of the three formulations given above.

\section{Gravitational field}

Now we write that part of the Lagrangian that describes the gravitational field.  I am now limiting this part to the Linear approximation to Einstein's full theory of General Relativity.

\begin{equation}
g_{\mu \nu} (x) = \eta_{\mu \nu} + h_{\mu \nu} (x) -\frac{1}{2} \eta _{\mu \nu} h + O(G^2) = \eta_{\mu \nu} + \lambda_{\mu \nu}(x) + O(G^2), \label{d1}
\end{equation}
where $h_{\mu \nu}$ is first order in G and $h = \eta^{\mu \nu} h_{\mu \nu}$; and henceforward we use the Minkowski metric $\eta_{\mu \nu}$ to raise and lower indices.

The equation of motion (Einstein's theory in the linear approximation) is,
\begin{eqnarray}
\partial^\alpha \partial_\alpha \; h_{\mu \nu}(x) = -16 \pi G T_{\mu \nu} (x),  \label{d2} \\
\partial^\alpha \partial_\alpha \; \lambda_{\mu \nu}(x) = -16 \pi G [T_{\mu \nu} (x) - \frac{1}{2} \eta_{\mu \nu}T(x)],  \label{d3}
\end{eqnarray}
with the gauge condition $ \partial^\mu h_{\mu \nu} =0$.

Let me try the following construction, which is first order in G:
\begin{equation}
{\cal{L}}_G = \frac{1}{64\pi G} [(\partial^\alpha \lambda^{\mu \nu}) (\partial_\alpha \lambda_{\mu \nu}) - \frac{1}{2} (\partial^\alpha \lambda) (\partial_\alpha \lambda)],  \label{d4}
\end{equation}
where  $\lambda = \eta^{\mu \nu} \lambda_{\mu \nu}$. When we go through variation of the action it will involve partial integrations over time and space. We assume that the deviations of the metric from the Minkowski form are contained in space, so there should be no surface terms from the partial integration over space. Regarding integration over time, the usual action rules say that there is no variation at the time endpoints, whatever they may be. 

\begin{equation}
\int d^4x \;\frac{\partial {\cal{L}}_G} {\partial g_{\mu \nu}(x)} = \int d^4x\;\frac{-1}{32 \pi G}[ \partial^\alpha \partial_\alpha \lambda^{\mu \nu}(x)  - \frac{1}{2}\eta^{\mu \nu} \partial^\alpha \partial_\alpha \lambda];  \label{d5}
\end{equation}

Using the matter Lagrangian density from the previous section we have,
\begin{equation}
\frac{\partial {\cal{L}}_M} {\partial g_{\mu \nu} (x)} = \sum_a  \frac{\zeta_a \epsilon_a}{2}m_a \int d\tau_a\;  \dot{\xi}_a^\mu \dot{\xi}_a ^\nu  \delta^4(x - \xi_a (\tau_a)) = \frac{-1}{2} T^{\mu \nu}(x). \label{c14}
\end{equation}
This defines the energy-momentum tensor $T^{\mu \nu} (x)$. Thus we have the complete Lagrangian density, ${\cal{L}} = {\cal{L}}_M + {\cal{L}}_G $ giving us the correct equations of motion (\ref{d3}) upon variation of the metric. Now we calculate the Hamiltonian for this entire system of particles and (linearized) gravitational field, adding (\ref{c12a})

\begin{equation}
H =H_M + H_G = - \sum_a p_{a\;0} + \frac{1}{64\pi G} \int d^3x [ (\partial_\alpha \lambda^{\mu \nu})( \partial_\alpha \lambda_{\mu \nu}) - \frac{1}{2} (\partial_\alpha \lambda)(\partial_\alpha\lambda)] ,\label{d6}
\end{equation}
where that second term has no longer the Lorentz invariant $\partial^\alpha ...  \partial_\alpha$ but what looks like a Euclidean sum.

Let's explore this result. For any set of fields $\varphi_b(x)$ in 3-dimensional Euclidean space that are produced by  localized source densities $\rho_b(x)$,  we have
\begin{eqnarray}
\triangle \varphi _b= \partial_i \partial_i \varphi_b = -4 \pi \rho_b, \;\; \varphi_b (x)= \int d^3 x' \frac{\rho_b(x')}{|\textbf{x} - \textbf{x}'|} \sim M_b/r ,  \label{d7} \\
\frac{1}{4 \pi} \int d^3 x \;(\partial_i \varphi_b)(\partial_i \varphi_c) =  \int d^3x\;\int d^3x'\; \frac{\rho_b(x)\;\rho_c(x')}{|\textbf{x} - \textbf{x}'|},  \label{d8}
\end{eqnarray}
where the $\sim$ means at a large distance $r$ from the source. Thus we have for the gravitational field part of the Hamiltonian, setting $\partial_t \lambda^{\mu \nu} = 0$,
\begin{equation}
H_G = G \int d^3x\;\int d^3 x'\; \frac{1}{|\textbf{x} - \textbf{x}'|} [T^{\mu \nu} (\textbf{x}) T_{\mu\nu} (\textbf{x}') - \frac{1}{2}T (\textbf{x}) T (\textbf{x}')] .  \label{d9}
\end{equation}

Putting in the earlier formula for the source T:
\begin{equation}
H_G = \sum_{a, b} \frac{G(\zeta_a \epsilon_a m_a \gamma_a)(\zeta_b \epsilon_b m_b \gamma_b)} {r_{ab}} \;[(1-\textbf{v}_a\cdot \textbf{v}_b)^2 - \frac{1}{2} (1-v_a^2)(1-v_b^2)].  \label{d10}
\end{equation}

For future use we have from (\ref{d3}) and (\ref{c14}),
\begin{equation}
\lambda^{\mu \nu} (\textbf{x}) = 4 G \sum_b \frac{\omega_b m_b \gamma_b}{|\textbf{x} -\textbf{x}_b|} [v_b^\mu v_b^\nu - \frac{1}{2}\eta^{\mu \nu}(1-v_b^2)],\label{d10a}
\end{equation}
where $\omega = \zeta \epsilon$ for each particle.

If I look at the non-relativistic limit, $v_a \rightarrow 0$, this term becomes
\begin{equation}
H_G \rightarrow \frac{1}{2}  \sum_a \sum_b \frac{Gm_a m_b}{r_{ab}},  \label{d11}
\end{equation}
which looks familiar. In the very high energy limit, $v_a \rightarrow 1$, we see,

\begin{equation}
H_G \rightarrow \sum_{a, b} \frac{G(\omega_a  m_a \gamma_a)(\omega_b  m_b \gamma_b)} {r_{ab}} \;(1-cos \theta_{ab})^2 ,  \label{d10b}
\end{equation}
where $\theta_{ab}$ is the angle between the two velocity vectors. 
For low energy tachyons we have another limit, $v_a \rightarrow \infty$
\begin{equation}
H_G \rightarrow  \sum_a \sum_b \frac{Gm_a m_b}{r_{ab}}\zeta_a \zeta_b v_a v_b (cos^2 \theta_{ab} - \frac{1}{2}).  \label{d12}
\end{equation}
This formula is not familiar; but one may want to compare it to the expression for the energy of the magnetic field produce by a spatial distribution of static electric currents. Note that the velocities may vary with spatial position of the particles, so this calculation is not as simple as it may look.

\section{Assembling the final formulas}

Now we will use the formulas (\ref{d10a}) for $\lambda_{\mu \nu}$ from the last section, insert them into the particle Hamiltonian formulas - three versions of that - from Section 5, and then add the Hamiltonian for the gravitational field itself (\ref{d10}).  For the Kinetic Energy I will use various definitions of the free particle energies: $E = m \gamma$, $E^* = \sqrt{p^i p^i +\epsilon m^2}$, or $E_* = \sqrt{p_ip_i +\epsilon m^2}$.  The Potential Energy terms are the real focus of interest, and we will write them in terms of the Energy E and the velocity v. As noted earlier, I write the $\pm 1$ factors as $\omega_a =\zeta_a \epsilon_a$, where the labels $a,b$ identify the individual particles.

First, using Formula 1, which is reliable for particles of low energy (and maybe moderate energy) of both species, ordinary ($\epsilon = +1$) and tachyon ($\epsilon = -1$), we get,
\begin{eqnarray}
\textbf{Formula 1:} \;\;\;\;\; H = H_M + H_G =- \sum_a \omega_a E_a - \sum_{a, b} \frac{G\; \omega_a E_a\;\omega_b E_b}{r_{ab}} Z_{ab}\;\;\;\;\;\;\;\;\;\;,\label{e1} \\
Z_{ab} = 2 -4 \textbf{v}_a \cdot \textbf{v}_b + v_a^2 + v_b^2 -[\epsilon_a \gamma_a^2 + \epsilon_b \gamma_b^2 +1] [(1- \textbf{v}_a \cdot \textbf{v}_b )^2 - (1/2)(1-v_a^2)(1-v_b^2)].\label{e2}
\end{eqnarray}
And here are several interesting sub-cases to be noted.

For all low energy ordinary particles  $v's \rightarrow 0$, $Z_{ab} \rightarrow 1/2$ and we have the Newtonian formula (\ref{z2}). We noted earlier that all three Formulations lead to this result.

For all low energy tachyons, $v's >>1$, 
\begin{equation}
Z_{ab} \rightarrow v_a^2 v_b^2 (1/2 - cos^2 \theta_{ab})\label{e3}
\end{equation}
where $\theta_{ab}$ is the angle between those two velocity vectors. It is noteworthy that this result comes entirely from $H_G$. In my first work \cite{CS1} I proposed that low energy tachyons could be attracted to one another in a rope-like structure. But this formula says NO to that model. Co-moving particles means that, on average, $\theta_{ab}$ is close to zero. Thus this $Z_{ab}$ is negative and the potential energy is seen as positive, increasing at small distances. We interpret that as a repulsive, not an attractive, force.

Another model would imagine a gas-like dispersion of the tachyons. With all directions of motion equally populated we have $<cos^2 \theta> = 1/3$ and $<Z_{ab}>$ is thus positive. This implies an attractive force between same-type tachyons, that is for $\zeta_a = \zeta_b$; but a repulsive force between opposite-type tachyons, $\zeta_a = - \zeta_b$. This suggest some very provocative physics about tachyon neutrinos in the cosmos. In reference \cite{CS4} I showed that the energy-momentum tensor for tachyon neutrinos should have a sign that changes with the helicity of the particles. That is what the factor $\zeta$ is for. 

Still more model possibilities need to be explored. That will be later work.

We can also use this formula (\ref{e2}) to look at low energy tachyons interacting with low energy ordinary particles. Here we are looking at one velocity $v_a$ going to zero and the other velocity $v_b$ going to infinity. There will be two contributions from the particles: and when we add (\ref{d10}) for the field, again two terms. The interaction potential has the form (\ref{e1}) with,
\begin{equation}
Z_{ab} \rightarrow 3 - 4 (v_a^2 v_b^2) (cos^2 \theta_{ab} - 1/2) \label{e4}.
\end{equation}
This is a rather weak potential: there are terms of order $v_b^2$ in both $H_M$ and $H_G$, but they cancel out.  The sign of $Z_{ab}$ depends on the details of angular dependence and the relative magnitudes of the two velocities.This may be interesting in looking at a region of evolution of the universe with temperatures far below the mass of ordinary matter but not yet far below the mass of tachyon-neutrinos. If we thus take $v_av_b << 1$, Then $Z_{ab}$ is positive. Alternatively, if we average over the angles, then $Z_{ab}$ is again positive regardless of the relative magnitude of velocities. But the sign of the overall interaction will depend on $\zeta_b$. If $\zeta_b = +1$, then we would call this an attractive force; but that is not the choice for $\zeta$ that gives us the explanation of Dark Energy.\cite{CS3} This may be relevant for a model of Dark Matter that posits a gas of tachyon neutrinos attracted to ordinary matter in galaxies.  More study is needed here.
 
Now we turn to Formulas 2 and 3 from Section 5. These should be reliable away from the region of low energy tachyons, i.e., E not near zero.
\begin{eqnarray}
\textbf{Formula 2}: \;\;\; H = H_M + H_G = -\sum_a \omega_a E_a^* - \sum \frac{G\omega_a E_a \omega_b E_b}{r_{ab}} Y_{ab}, \\
Y_{ab} = 1/2  - (v_a^2+v_b^2)/2 - 3 v_a^2v_b^2 (cos^2\theta_{ab}-1/2).
\end{eqnarray}
\begin{eqnarray}
\textbf{Formula 3}: \;\;\; H = H_M + H_G =- \sum_a \omega_a (E_*)_a - \sum \frac{G\omega_a E_a \omega_b E_b}{r_{ab}} X_{ab}, \\
X_{ab} = 1/2 + (v_a^2+v_b^2)/2 + v_a^2v_b^2 (cos^2\theta_{ab}-1/2).
\end{eqnarray}
Here I have dropped terms $\textbf{v}_a \cdot\textbf{v}_b$ from $H_G$ to be consistent with the simplification $g_{0i} = 0$ used for Formulas 2 and 3.

Averaging over angles $<Y_{ab}> = (1-v_a^2)(1-v_b^2)/2$, $<X_{ab}> = (1+v_a^2 + v_b^2 - v_a^2 v_b^2/3)/2$.
How these later two formulas may be useful is for later study. For high energy particles, $v \rightarrow 1$, both of these formulas give positive values, which is reassuring. The formula for $<Y>$ looks most reasonable, with the factors $(1-v^2)$ damping down the strength of the interaction as one goes to high energies. On the other hand, Formula 3 is the one that adheres to the canonical formalism.

\section{Conclusions}

This paper has gone through three different formulations. I should explain this.  

The initial motive was to find some formulation of the gravitational dynamics for a large collection of tachyons, particularly at low energies, $E << m$, where we expected they would produce unusually strong gravitational fields that would have significant physical effects observable in cosmological studies. Constructing a Hamiltonian for such a system, many relativistic particles plus their gravitational field, seemed like the way to go. 

In nonrelativistic dynamics we are used to writing $H = K.E. + P.E.$ with the first term depending only on v (or p, which is a function of v), while the second term depends only on x. We know how to "read" the P.E. and say whether the force will be attractive or repulsive. An attractive force might or might not lead to bound states; a repulsive force would surely allow only scattering states. That is what I had hoped to do here, with tachyons and with Einstein's General Relativity. But this - even with our special assumption of a static metric - turns out to involve a much messier mixing of spatial coordinates and velocities. This led to my earlier blunders, noted in Section 5, which I was directed to restudy upon receiving sharp questioning from this journal's (anonymous) reviewer. 

In this third version I explore three different approaches and serve up three different Formulas, with the following strict rules: Formula 1 is best used only fo low energy particles. Formulas 2 and 3 should not be used for low energy tachyons.

With the general formulas and various specialized forms, as presented in the last Section, one may start the hard work of trying to build models of tachyon flows that are stable and confined and may contribute effectively to the proposition that neutrinos-as-tachyons may explain the observed phenomena now ascribed to Dark Matter. There are also provocative hints about how a gas of tachyons may interact gravitationally; and this will be especially worth studying if the search for confined tachyons should fail.

The one sharp conclusion from this work is the \emph{rejection} of my original \cite{CS1} model of low energy tachyons being attracted to one another in a rope-like structure. 

I will save my own further modeling for a separate paper, while offering the above mathematical tools for others to explore independently. 

As a bit of self-criticism, I offer the thought that the Static model, upon which this paper is based, may be very questionable for the high energy situations considered in Cosmology, although I have blithely ventured into that domain above. Perhaps, as an alternative to the assumption of "static  flows", one may be able to invoke the idea of "time averaging" in order to justify using a static metric.
Also, in Section 6 I did some partial integrations and ignored any surface terms on the assumption of a localized source; but then in Section 7 I also considered an extensive gas of particles as the source. Perhaps this is fixable.

Overall, I am disappointed that I could not find a single formula for the Hamiltonian that is valid for all energies of the particles. Nevertheless, what has been presented here should be useful tools as one goes on to explore specific models.

\vskip 1cm
{\bf Acknowledgments}
\newline
I am most grateful to Korkut Bardakci for many helpful discussions. I also want to thank the (anonymous) reviewer whose sharp questioning led me to uncover, and to fix, significant faults in the first two versions of this paper.\cite{CS5}

\vskip 1cm
\setcounter{equation}{0}
\def\theequation{A.\arabic{equation}}
\boldmath
\noindent{\bf Appendix A: Debunking the anti-tachyon myths}
\unboldmath
\vskip 0.5cm

Some people believe that the Earth is flat; and they can see this is true with their own eyes. Physicists are convinced otherwise; and they can cite abundant evidence on their side.

However, when it comes to Tachyons (faster-than-light particles), a great many physicists believe that they do not, and some will say that they can not, exist; and a number of reasons are recited in support of that prejudice.

In my several papers exploring mathematical frameworks for how Tachyons might fit into physical theory and experiments I have taken the trouble to lay out careful reasoning to debunk those prejudices.  Do I have to review all those arguments in every new paper I write? Maybe. 

Firstly, all my work is done strictly within the established mathematical frameworks of Special and General Relativity. (Some other authors have violated those bounds.)

In my 2011 JMP paper, Appendix A looks at a scenario of sending tachyon signals between earth and a distant rocket ship, alleging \emph{a causal paradox}. It is argued that an exchange of tachyon signals can lead to a response arriving before the original message was sent out. Simply replacing the point particle by a wave packet shows that, when one carries out the relevant Lorentz transformation, the distinction between sending (emitting) and receiving (absorbing) a tachyon can disappear.

In my 2016 IJMPA paper, I state the appropriate \emph{principle of causality} for tachyons - no propagation slower than the speed of light; and this leads to a consistent mathematical formalism for quantizing such fields. This provides an alternative to the canonical formalism, which is wrong for tachyon fields.

In my 2018 IJMPA paper, Section 2 examines the role of tachyons engaged in a general multi-particle interaction. The common idea that \emph{negative energy states} imply physical instability of the system is debunked by recognizing that the naming of $in$ and $out$ states is not Lorentz invariant. The total energy and momentum are properly conserved.

In my 2016 paper on quantizing tachyon fields, especially for the spin 1/2 (Dirac) case, I deal with the \emph{Little Group O(2,1)} by introducing an indefinite metric (the helicity) into the Fock space.

Then there are experiments, a number of which over the years have claimed to observe neutrinos as tachyons, and then been revised to the opposite conclusion. The 2011 \emph{OPERA experiment} looked at 20 GeV neutrinos and first reported that they travelled faster than light by 1 part in 40,000. That would imply a tachyon mass of about 100 MeV. But we know that neutrino mass is around 0.1 eV; and the excess velocity (v-c) goes as the square of the mass-to-energy ratio. That puts us 14 orders of magnitude below the original (wrong) observation.

Finally, there are theoretical efforts to derive the existence of known particles from some abstract field with complicated self-interactions. The simplest model is a scalar field with a potential that looks like W. If one expands around the central peak, then the resulting particles are found to be tachyons (negative mass-squared). But then one recognizes that those states are unstable; one should instead expand about the minima of W, where one gets ordinary particles.  I am not involved in that sort of theorizing. 

I start with the question: If tachyons do exist, how would we describe them within our customary mathematical frameworks? The starting point is the relativistically invariant form for any 4-vector (e.g., the energy-momentum of a particle): \newline $p^\mu p_\mu = constant$. That constant may be positive, zero, or negative.

\end{document}